\documentstyle[12pt,epsfig]{article}

\def\gappeq{\mathrel{\rlap {\raise.5ex\hbox{$>$}}
{\lower.5ex\hbox{$\sim$}}}}

\def\lappeq{\mathrel{\rlap{\raise.5ex\hbox{$<$}}
{\lower.5ex\hbox{$\sim$}}}}

\begin{document}

\newcommand{\be}{\begin{equation}}
\newcommand{\ee}{\end{equation}}
\newcommand{\beq}{\begin{eqnarray}}
\newcommand{\eeq}{\end{eqnarray}}
\newcommand{\la}{\langle}
\newcommand{\ra}{\rangle}

\pagestyle{empty}
\begin{flushright}
CERN-TH/2000-075 \\
hep-ph/0003083 \\
\end{flushright}
\vspace*{5mm}
\begin{center}
{\bf DOES EFFICIENCY OF HIGH ENERGY COLLISIONS \\
DEPEND ON A HARD SCALE?} \\
\vspace*{1cm} 
{\bf A.V. Kisselev}$^{*}$ \\
\vspace{0.3cm}
Theoretical Physics Division, CERN \\
CH - 1211 Geneva 23 \\
\vspace{0.3cm}
and \\
\vspace{0.3cm}
{\bf V.A. Petrov}$^{\dag}$ \\
\vspace{0.3cm}
Institute for High Energy Physics, \\
142284 Protvino, Russia \\
\vspace*{2cm}  
{\bf ABSTRACT} \\ 
\end{center}
\vspace*{5mm}
\noindent
The multiplicity of charged hadrons in the current fragmentation region of both
the c.m.s. and the Breit frame of deep inelastic scattering is calculated  and
compared with the HERA data. The results are in agreeement with Yang's
hypothesis that the efficiency of high energy processes increases at  larger
momentum transfer. 
\vspace*{3cm} 

\begin{flushleft} CERN-TH/2000-075 \\
March 2000
\end{flushleft}


\noindent 
\rule[.1in]{16.5cm}{.002in}

\noindent
{\small $^{*)}$ Permanent address: Institute for High Energy Physics,
142284 Protvino, Russia \\
\hphantom{*)}E-mail: kisselev@mx.ihep.su \\
$^{\dag)}$ E-mail: petrov@mx.ihep.su}

\vfill\eject

\setcounter{page}{1}
\pagestyle{plain}

\section{Introduction}

It seems quite natural to expect that the harder a high-energy collision is,
the higher is the number of fragments. One of the most tractable and widely
explored processes where this phenomenon can be seen is deeply inelastic
scattering (DIS) with a possibility to change a hard scale ($Q^2$, gauge boson
virtuality) and to detect its influence (if any) on the hadronic invariant
mass ($W$) ''efficiency''. The simplest measure of this efficiency is the
multiplicity of secondaries.

In 1969 Yang and his collaborators~\cite{Yang}, basing themselves on the 
``fragmentation picture`` of violent collisions, made a qualitative prediction:
``...for larger values of the momentum transfer $t$, the breakup process favors
larger multiplicities of hadrons``  (at fixed hadronic mass). Early searches for
this effect were inconclusive in both theory and experiment~\cite{Gibbard}.

In the framework of QCD a quantitative result has been obtained in
Refs.~\cite{Dzhaparidze,KisselevY0}: it appears that QCD gluon bremsstrahlung
leads to an increase of the hadron multiplicities in DIS, with an increase  of
$Q^2$ at fixed hadronic mass $W$, but that this increase is very slow. The
distinctive feature of this result is that $\la n \ra^{DIS}(W,Q^2)$ has a
finite limit at $Q^2 \rightarrow \infty$ and $W$ fixed.

Later, another result has been claimed in Ref.~\cite{Gribov}, which predicted
an infinite and quite rapid growth of $\la  n \ra^{DIS}(W,Q^2)$ with $Q^2$. 
In the course of the inference of this result it was supposed that the influence 
of the (non-perturbative) composite structure of the nucleon is negligible
while in~\cite{Dzhaparidze,KisselevY0} it plays a key role in the slowness of
the $Q^2$ dependence of the multiplicity at fixed $W$. 
There is even more trivial objection. On general grounds, the infinite growth
with $Q^2$ at fixed $W$ is impossible because of the {\em apriori} kinematical
bound
\be
\la n \ra^{DIS}(W,Q^2) \leq \frac{W}{m}, \label{1}
\ee
where $m$ is some effective mass.

Quite recently, a  weak dependence on $Q^2$ in the framework of the Dual Parton 
Model was mentioned in~\cite{Capella}.

Experimentally a statistically significant effect of the slow growth of $\la  n
\ra^{DIS}(W,Q^2)$ was established in $\nu (\bar \nu) p$
interactions~\cite{WA25}  and in $\mu^+ p$ interactions~\cite{EMC}. The results
of the EMC~\cite{EMC} have been described in the framework of QCD in
Ref.~\cite{PetrovEMC}. 

However, subsequent measurements at HERA (H1) were interpreted as a practical
$Q^2$ independence~\cite{H1CMS} of $\la  n \ra^{DIS}(W,Q^2)$ (for the current
hemisphere in the hadronic c.m.s.), while H1~\cite{H195,H197} and
ZEUS~\cite{ZEUS95,ZEUS99} reported quite fast $Q^2$ dependence for the current
hemisphere in the Breit frame. It should be noted, however, that this last
result concerns different bins in $W$ for changing $Q^2$ values. Anyway, the
situation is  controversial and therefore very interesting.

In this paper we give our own interpretation of the HERA data on charged hadron
multiplicities in the current fragmentation region; as will be seen in the text
below, these are in agreement with Yang's general hypothesis~\cite{Yang} and 
our early QCD results~\cite{Dzhaparidze,KisselevY0} (see also the
review~\cite{KisselevRev}).

\section{Hadronic Spectrum and Multiplicity in DIS}

According to the factorization for inclusive spectra in DIS, the hadronic 
spectrum in DIS is represented by two terms: 
\be
\frac{dn}{dy}^{DIS} \! \! \! (W,Q^2) = \int_{x_0}^1 \!\frac{dz}{z} 
w(x,z,Q^2) \frac{d \hat{n}}{dy}(W_{eff},Q^2) + \frac{dn_0}{dy}, \label{2}
\ee
where $x_0 = x + (1 - x)(m_h/W) \exp(-y)$, $y$ is the rapidity of the detected
hadron and $m_h$ is its mass. 
In Eq.~(\ref{2}) $d \hat{n}/dy$ defines the hadronic spectrum in partonic
subprocess, while the quantity $dn_0/dy$  describes the spectrum of the proton
remnant. The latter does not contribute to the  current fragmentation region 
of DIS at HERA energies. 

Correspondingly, the average hadronic multiplicity in DIS is represented by
\be
\la n \ra^{DIS}(W,Q^2) = \int_{x_0}^1 \!\frac{dz}{z} 
w(x,z,Q^2) \la \hat{n} \ra (W_{eff},Q^2) + \la n_0 \ra. \label{3}
\ee

For small $x$ the weight $w(x,z,Q^2)$ in (\ref{2}), (\ref{3}) is of the form:
\beq
w(x,z,Q^2) &=& D_g^q \left(\frac{x}{z},Q^2,Q_0^2 \right) 
f_g(z,Q_0^2) \nonumber \\
&\times& \left( \int_{x_0}^1 \!\frac{dz}{z} 
D_g^q \left(\frac{x}{z},Q^2,Q_0^2 \right) f_g(z,Q_0^2) \right)^{-1}.
\label{4}
\eeq
As can be seen, the hadronic spectrum in partonic subprocess, $d \hat{n}/dy$,
and the hadronic multiplicity $\la \hat{n} \ra$ depend on the effective energy, 
which is smaller than $W$:
\be
W_{eff}^2 = \frac{z - x}{1 - x} \, W^2. \label{6}
\ee

In what follows, we shall work in the c.m.s. of the final hadrons.
In terms of rapidity, the current region in the c.m.s. corresponds to 
\be
-Y < y < 0 \label{8}
\ee
(it is assumed that the proton goes in the positive direction).

In our papers~\cite{Dzhaparidze,KisselevY0} it has been established that
the total hadronic multiplicity in the partonic subprocess of DIS is related to 
the hadronic multiplicity in $e^+e^-$ annihilation:
\be
\la \hat{n} \ra (W,Q^2) \simeq \la n \ra^{e^+e^-} (W) \label{10}
\ee
(up to small NLO corrections, which decrease in $Q^2$).

In the partonic subprocess, the rapidity varies in the range
\be
- \hat{Y} < y + y_0 < \hat{Y}, \label{12}
\ee
where $\hat{Y} = \ln (W_{eff}/m_h)$ and 
\be
y_0 = \frac{1}{2} \ln \left( \frac{1 - x}{1 - z} \right).
\label{14}
\ee The quantity $y_0$ determines the rapidity of the centre of mass of the
partonic subprocess in the centre of mass of the complete process. On
integration over $z$, the region~(\ref{12}) is "smeared" into the region
\be
-Y < y < Y, \label{16}
\ee
with $ Y = \ln (W/m_h)$.

The average value of the effective energy in (\ref{2}), (\ref{3}), available for 
particle production, appeared to be dependent on both $W$ and 
$Q^2$~\cite{Dzhaparidze,KisselevY0}:
\be
\la W_{eff} \ra^2 \simeq   \kappa (Q^2) W^2. \label{18}
\ee
The efficiency factor $\kappa(Q^2)$, which stands in front of $W^2$ in 
(\ref{18}), is much less than 1 and grows slowly in $Q^2$.

From formulas (\ref{3}), (\ref{10}) and (\ref{18}), one can see that the rise 
of the average hadronic multiplicity in DIS has the same physical nature as in 
$e^+e^-$ annihilation. For the first time this behaviour has been experimentally
established by H1 in 1996~\cite{H1CMS}.

However, the QCD growth of $\la \hat{n} \ra$ is 
delayed in DIS by the bound-state effects and the slow QCD evolution of the
structure function. This is why we predicted that the $Q^2$ dependence of 
$\la \hat{n} \ra$ at fixed $W$ should remain numerically weak at HERA
energies~\cite{PetrovEMC,KisselevRev}.

It follows from (\ref{12}), (\ref{14}) that the centre of the spectrum is
shifted to the region of positive rapidities and tends to zero at 
asymptotically high $Q^2$~\cite{KisselevY0}:
\be
\la y_0 \ra \left|_{Q^2 \rightarrow \infty} \sim 
\frac{1}{\ln(\ln Q^2))}. \right. \label{20}
\ee

The hadronic spectrum in partonic subprocess in the c.m.s. of DIS has the form
\be
\frac{d \hat{n}^h}{dy} = n^{e^+e^-}(W_{eff},Q^2) \,
\bar D^h(W_{eff},y). \label{22}
\ee
Normalization in the RHS of Eq.~(\ref{22}) is done
 in agreement with
formula~(\ref{3}).

\section{Hadronic Multiplicities in the Current \newline
         Fragmentation Region}

To calculate the multiplicity of charged hadrons in the current fragmentation 
region, we have to define expressions of the quark distribution at
small $x$, of the hadronic spectrum in the partonic subprocess as well as of the 
multiplicity of charged  hadrons in $e^+e^-$ annihilation.

For the quark distribution, we use an analytical expression
from Ref.~\cite{Kotikov}, in the case of soft initial conditions.
As was shown in \cite{Kotikov}, at small $x$ it is in good agreement with the 
data on the structure function from HERA  in the wide ranges of $Q^2$. Namely,
at high $Q^2$, we have 
\be
D^q_g(z,Q^2) \sim  r I_1(t)\exp(-d \xi /2), \label{24}
\ee
with $d = \beta_0 + 20 N_f/27$.
The variable
\be
t= 2 \sqrt{6 \xi \ln \left(\frac{1}{z} \right)} \label{26}
\ee
is related to the QCD evolution parameter
\be
\xi = \frac{2}{\beta_0} \ln \left(\frac{\alpha(Q_0^2)}{\alpha(Q^2)} 
\right), \label{28}
\ee 
where $\beta_0 = 11 - 2 N_f/3$ is the  $\beta$-function in lowest order and 
\be
r = \frac{t}{2 \ln(1/z)}. \label{30}
\ee 
The quark and gluon distributions from Ref.~\cite{Kotikov} obey the GLAP 
evolution equations~\cite{GLAP}.

The expression of the initial gluon distribution at $z$ closed
to $1$ is chosen to have the following form
\be
f_g(z,Q_0^2) \left|_{z \rightarrow 1} \sim (1 - z)^{n_g}. \right. \label{32}
\ee
We have omitted constant factors in the RHS of Eqs.~(\ref{24}) and (\ref{32})
as they do not influence our final results.

The spectrum of hadrons in the partonic process $\bar D^h$ was calculated by 
many authors. We use the expression from Refs.~\cite{Fong} ($N$ is a
normalization factor):
\be
\bar D^h(W,\zeta) = \frac{N}{\sigma \sqrt{2 \pi}} \exp \left[ 
\frac{1}{8}k - \frac{1}{2}s \delta - \frac{1}{4} (2+k) \delta^2 
+ \frac{1}{6}s \delta^3 + \frac{1}{24}k\delta^4 \right] \label{34}
\ee
calculated in the variable
\be
\zeta = \ln \left(\frac{W}{E_h} \right). \label{36}
\ee
Here $E_h$ is the energy of the detected hadron.

The average value of $\zeta$, $\zeta_0$, and its dispersion $\sigma$
are given by the formulas:
\be
\zeta_0 =  \frac{1}{2} \tau \left(1 + \frac{\rho}{24}
\sqrt{\frac{48}{\beta_0 \tau}} \right) \left(1 - 
\frac{\omega}{6 \tau} \right), \label{40}
\ee
\be
\sigma = \sqrt{\frac{\tau}{3}} \left( \frac{\beta_0 \tau}{48}
\right)^{1/4}
\left( 1 - \frac{\beta_0}{64} \sqrt{\frac{48}{\beta_0 \tau}}
\right)
\left(1 + \frac{\omega}{8 \tau} \right), \label{42}
\ee 
where
\be
\tau = \ln \left(\frac{W}{\Lambda} \right) \label{44}
\ee
and 
\be
s = - \frac{\rho}{16}\sqrt{\frac{3}{\tau}} \left( \frac{48}{\beta_0 
\tau} \right)^{1/4} \left(1 + \frac{\omega}{4 \tau} \right), 
\label{46}
\ee 
\be
k = - \frac{27}{5 \tau} \left( \sqrt{\frac{\beta_0 \tau}{48}} 
- \frac{\beta_0}{24} \right) \left(1 + \frac{5 \omega}{12 \tau} 
\right), \label{48}
\ee  
\be
\delta = \frac{\zeta - \zeta_0}{\sigma}. \label{50}
\ee
Here $\rho = 11 + 2 N_f/27$, $\omega = 1 + N_f/27$.

At low (effective) energies we use the fit of the  low-energy data on
multiplicity of charged hadrons in $e^+e^-$ annihilation from 
Ref.~\cite{LowW}:
\be
\la n \ra^{e^+e^-} = 2.67 + 0.48 \ln W^2, \label{52}
\ee
while for high energies ($W_{eff} > 10$ GeV) we apply the fit from 
Ref.~\cite{HighW}, which well decsribes $e^+e^-$ data up to LEP energies:
\be
\la n \ra^{e^+e^-} = -1.66 + 0.866 \exp (1.047 \sqrt{\ln W^2}). \label{54}
\ee
We have corrected (\ref{52}) for a fraction of the charged particles from 
$K^0_s$ and $\Lambda \, (\bar \Lambda)$ decays. 

Figure~1 represents the result of calculations of charged multiplicity in 
current hemisphere of the c.m.s. by using formulas~(\ref{2}) and (\ref{24})
(solid  curves) in comparison with the H1 data from Ref.~\cite{H1CMS}. As can
be seen, our QCD predictions  are in very good agreement with the data. These
are quite compatible with a slow growth of $\la n \ra^{DIS}(W,Q^2)$ in $Q^2$ 
at fixed $W$. 

Figure~2 demonstrates a rapid rise of $\la n \ra^{DIS}(W,Q^2)$ in the variable $W$
for different values of $Q^2$, which was predicted many years ago in 
Refs.~\cite{Dzhaparidze,KisselevY0} and seen previously in $e^+e^-$ 
annihilation (the very values of $Q^2$ taken from \cite{H1CMS}). Let us note
that the H1 data presented in Fig.~2  (see Table~4 in \cite{H1CMS}) do not 
correspond to some fixed values of $Q^2$, in contrast with the experimental 
points in Fig.~1.

In order to obtain  multiplicity of charged hadrons in current region of the
Breit frame,  the c.m.s. spectrum~(\ref{22}) must be integrated in the region
\be
-Y < y < y_B, \label{56}
\ee
where
\be
y_B = - \frac{1}{2} \ln \left( \frac{1 + v}{1 - v} \right)
\simeq - \frac{1}{2} \ln \left( \frac{1}{x} \right). \label{58}
\ee
The quantity
\be
v = \sqrt{1 - 4x(1 - x)} \label{60}
\ee
in the RHS of Eq.~(\ref{58}) is the velocity of the Breit frame in the c.m.s.
So, $y_B$ corresponds to zero rapidity in this frame. 

The results of our calculations of the multiplicity of charged hadrons 
in the current region of the Breit frame are presented in Fig.~3 as 
a function of $Q^2$, in comparison with the H1 data~\cite{H197} (solid squares)
and ZEUS data~\cite{ZEUS99} (solid circles).

The theoretical curves in Figs.~1--3 correspond to the following values of
parameters:
\be
Q_0^2 = 1 \mbox{ \rm GeV}^2, \quad \Lambda = 0.25 \mbox{ \rm GeV}.
\label{62}
\ee
The parameter $n_g = 6.1$ in Eq.~(\ref{34}) is taken from one of the MRST  
sets of parton distributions~\cite{MRST}.

It should be noted that the strong $Q^2$ dependence seen by H1 and ZEUS in the 
Breit frame has nothing to do with the $Q^2$ dependence of  $\la n
\ra^{DIS}(W,Q^2)$ in the c.m.s.  (see Fig.~1); to a large  extent it has a
kinematical origin.  The point is that an increase of $Q^2$  at fixed $W$ is
equivalent to an increase of $x$. As a result, the current
region of the Breit  frame~(\ref{56}) enlarges. Thus, the rapid growth of
hadronic multiplicity in the Breit frame in $Q^2$ (at fixed W) reflects a
strong increase of the hadronic spectrum towards the central region.  

As for the increase of $\la n \ra^{DIS}(x,Q^2)$ in $Q^2$ at fixed $x$ in the
Breit frame, it has been found that it is similar to that in $e^+e^-$
annihilation at high  $Q^2$, while there is a discrepancy between DIS and 
$e^+e^-$ data at low $Q^2$~\cite{H195}--\cite{ZEUS99}. Within our approach, it
can be understood as follows.  On the one hand, the increase of  $Q^2$ results
in an increase of the  height of the spectrum (because $W$ grows). On the other
hand, the position of the spectrum, $\la y_0 \ra$ as defined in  (\ref{14}),
tends towards the region of positive rapidities.

These two phenomena go in opposite directions. At $Q^2$ high enough, $\la y_0
\ra$ varies very slowly with  $Q^2$~(\ref{20}). As a result, the rise of
hadronic multiplicity in the Breit frame is analogous to that in $e^+e^-$
annihilation. At low $Q^2$, $\la y_0 \ra$ changes more 
significantly~\cite{KisselevY0}. This effect partially compensates the growth
of the spectrum and there appears a significant difference between DIS and 
$e^+e^-$ data.

Finally, it is interesting to analyse the case when $x$ increases while $Q^2$
remains fixed. In this, $W$ decreases, which results in a rapid decrease of the
spectrum. At the same time, however, the current region in the Breit frame 
becomes larger in accordance with formulas ~(\ref{56}), (\ref{58}). The effects are
of the same order but opposite in sign. The ZEUS data in the Breit frame  (see
Table~2 in Ref.~\cite{ZEUS99}) show that there is a slow  rise  of
the  hadronic multiplicity in the current hemisphere in the variable $x$ at 
different fixed values of $Q^2$.

\section{Conclusions}

In this paper we have presented the results of QCD calculations of the 
multiplicity of charged hadrons in the current hemisphere of DIS. Both the
c.m.s. and the Breit frame are considered. We have shown that the H1 data are
in agreement with Yang's hypothesis and our QCD predictions. Namely, the
efficiency of high energy collisions does weakly depend on the hard scale of
the process (momentum transfer $Q^2$). It means that {\it at fixed energy} the 
efficiency of a particle production in hard processes increases with the 
shrinking of the interaction region ($\sim 1/Q$, in DIS).

The observed rise of the hadronic multiplicity with $Q^2$  in the current
region of the Breit frame has both a dynamical and kinematical origin. This is
why a direct comparison of available DIS data in the Breit frame with $e^+e^-$
data is not completely correct.

New data from HERA on the hadronic multiplicity in the current region of the
c.m.s. as well as measurements of the total multiplicity in DIS as functions
of  two variables ($Q^2$ and $W$/$x$) would be very important.

\section*{Acknowledgements}
We would like to thank Professors G.~Altarelli, S.~Catani, A.B.~Kaidalov, 
M.~Mangano and G.~Veneziano for useful discussions. One of us (A.V.K.) is
indebted to Professors G.~Altarelli, S.~Catani and M.~Mangano  for their
support. We also thank Professor L.M.~Shcheglova for stimulating discussions 
of the HERA data and Professor A.L.~Kataev for sending us information on 
parton distributions.

\eject 

%
 \begin{figure}[htpb]
   \begin{center}
    \mbox{\epsfig{file=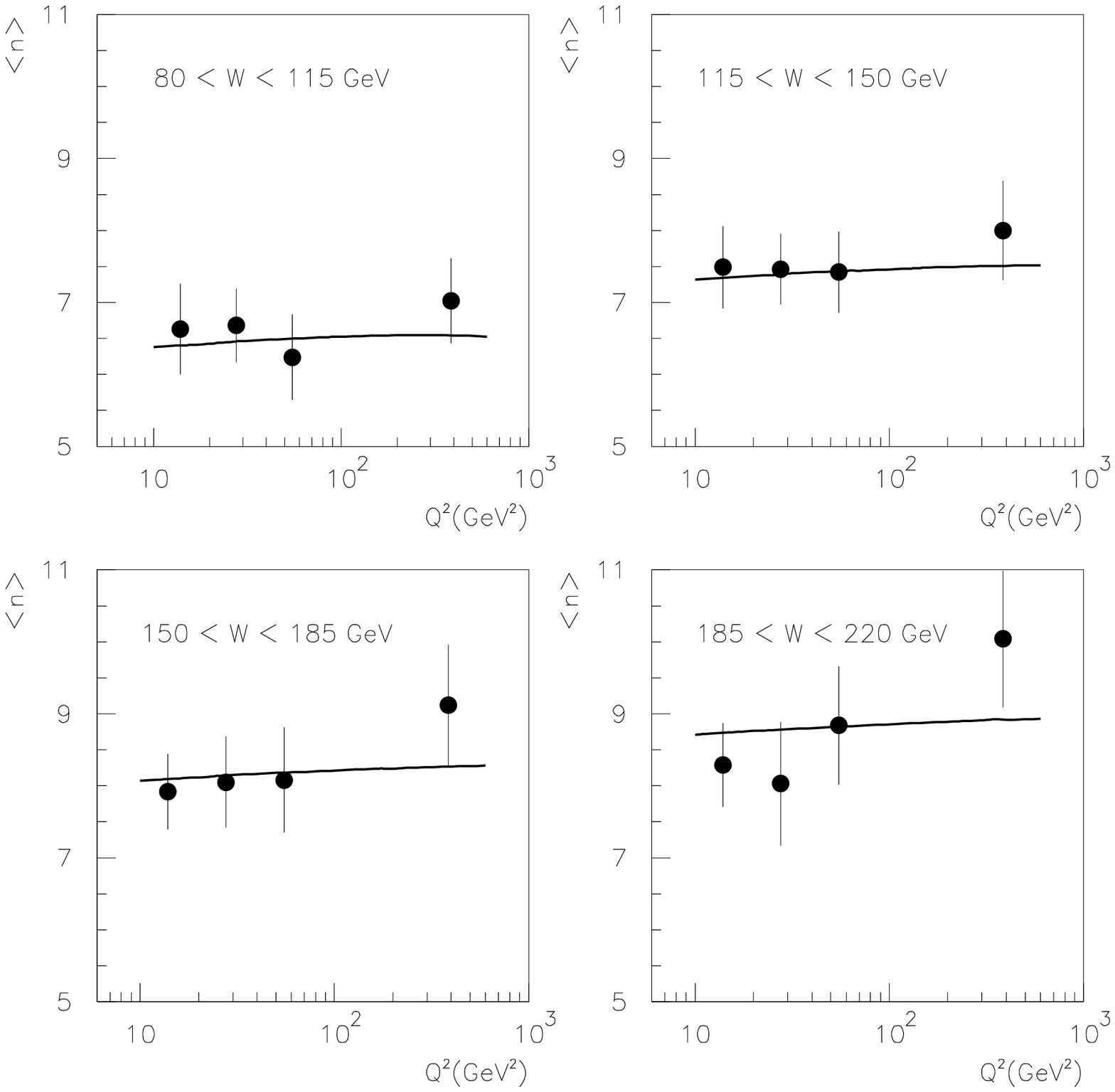,height=14cm,width=14cm}}
   \end{center}
   \caption{The $Q^2$ dependence of the multiplicity of charged  hadrons
   in the current fragmentation region of the c.m.s. in intervals of $W$}
  \end{figure} 
  
 \begin{figure}[htpb]
   \begin{center}
    \mbox{\epsfig{file=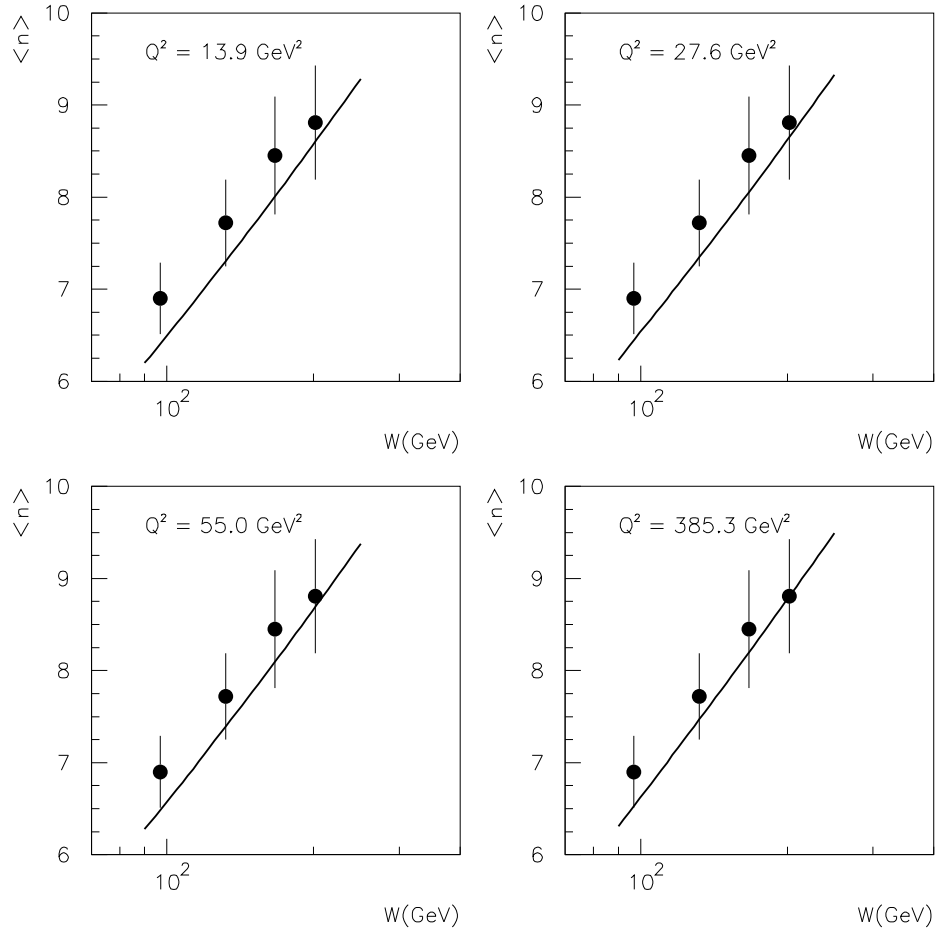,height=14cm,width=14cm}}
   \end{center}
   \caption{The $W$ dependence of the multiplicity of charged hadrons
   in the current fragmentation region of the c.m.s. at fixed values of 
   $Q^2$}
  \end{figure} 
  
 \begin{figure}[htpb]
   \begin{center}
    \mbox{\epsfig{file=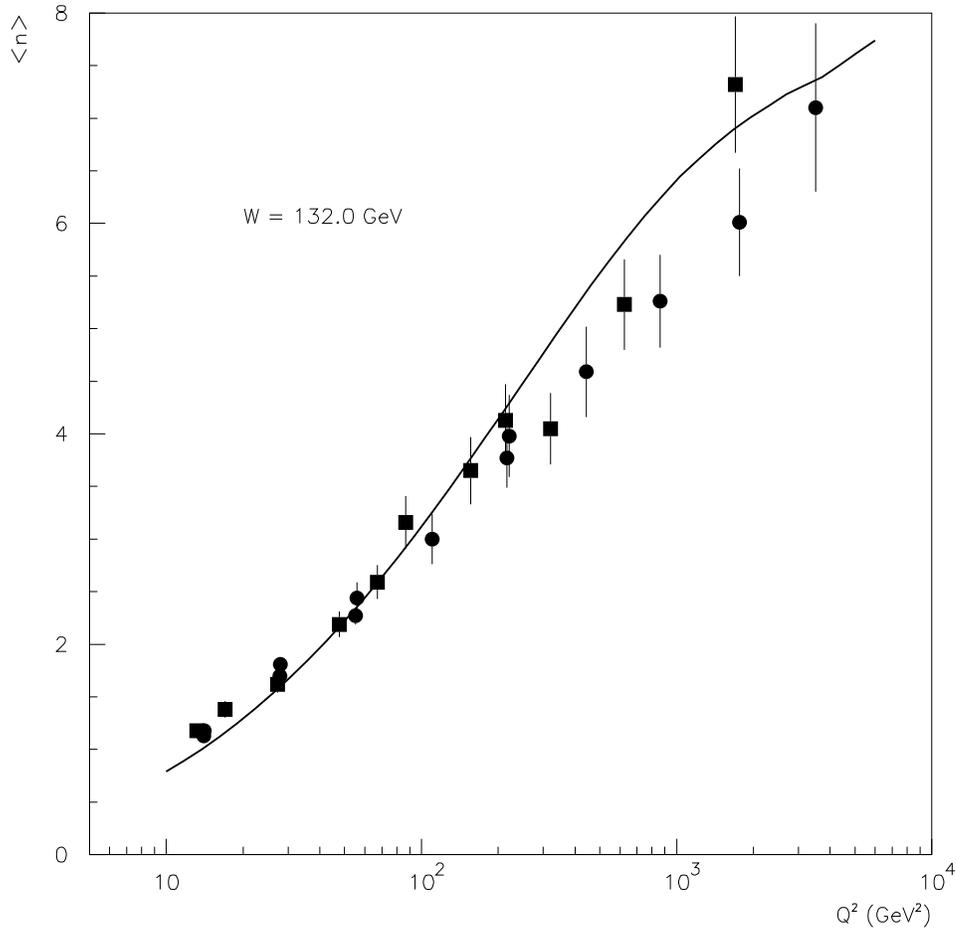,height=14cm,width=14cm}}
   \end{center}
   \caption{The $Q^2$ dependence of the multiplicity of charged hadrons
   in the current fragmentation region of the Breit frame at fixed values 
   of $W$}
  \end{figure} 
    
\end{document}